\documentclass[aps,pra,showpacs,10pt]{revtex4-1}
\usepackage{graphicx}
\usepackage{amsmath}

\input{epsf}
\begin{document}
\epsfverbosetrue
\def\la{{\langle}}
\def\ra{{\rangle}}
\def\vep{{\varepsilon}}
\newcommand{\beq}{\begin{equation}}
\newcommand{\eeq}{\end{equation}}
\newcommand{\beqa}{\begin{eqnarray}}
\newcommand{\eeqa}{\end{eqnarray}}
\newcommand{\da}{^\dagger}
\newcommand{\wh}{\widehat}
\newcommand{\os}[1]{\hat{#1}_{\hbox{\scriptsize {osc}}}}
\newcommand{\cn}[1]{\hat{#1}_{\hbox{\scriptsize{con}}}}
\newcommand{\sy}[1]{\hat{#1}_{\hbox{\scriptsize{sys}}}}

\draft

%\jobname
\title{Quantum decoherence of an anharmonic oscillator monitored by a Bose-Einstein condensate}
\author{D. Alonso}
\email{dalonso@ull.es}
\affiliation{Instituto Universitario de Estudios Avanzados (IUdEA)}
\affiliation{Departamento de F\'{\i}sica, 
Universidad de La Laguna, La Laguna E38204, Tenerife, Spain.}
\author{S. Brouard}
\email{sbrouard@ull.es}
\affiliation{Instituto Universitario de Estudios Avanzados (IUdEA)}
\affiliation{Departamento de F\'{\i}sica, 
Universidad de La Laguna, La Laguna E38204, Tenerife, Spain.}
\author{D. Sokolovski}
\email{dgsokol15@gmail.com}
\affiliation{Department of Physics Chemistry, University of the Basque Country, Leioa, Spain}
\affiliation{IKERBASQUE, Basque Foundation for Science, 48011, Bilbao, Spain.}
\begin{abstract}
The dynamics of a quantum anharmonic oscillator whose position is monitored by a Bose-Einstein condensate
(BEC) trapped in a symmetric double well potential is studied. The (non-exponential) decoherence induced
on the oscillator by the measuring device is analysed. A detailed quasiclassical and quantum analysis is
presented. In the first case, for an arbitrary initial coherent state, two different decoherence regimes
are observed: An initial Gaussian decay followed by a power law decay for longer times. The characteristic
time scales of both regimes are reported. Analytical approximated expressions are obtained in the full
quantum case where algebraic time decay of decoherence is observed.
%The decoherence process affecting an initial superposition is studied numerically in the quantum case.
\end{abstract}
\date{\today}
\pacs{03.65.Yz, 03.67.Ta, 03.75.Gg}
\maketitle
\vskip0.5cm

\section{Introduction}
Recent experiments \cite{hunger20101,treutlein20121} show that a great degree of coherent control is possible
between micromechanical oscillators and a Bose-Einstein condensate (BEC) of magnetically trapped Rubidium-87
atoms \cite{reichel20001}. Chip-based magneto traps offer a high degree of control when the BEC and a
micromechanical cantilever are brought close to distances of the order of the micrometer. At that
level of proximity between a cantilever and a BEC, the surface forces start to play a role. Such forces
allow one to couple coherently the collective dynamics of a condensate and a mechanical oscillator.
Accordingly, it is possible to study the interaction between trapped atoms and on-chip-solid-state systems
such as nano-micro mechanical oscillators \cite{schwab2005a,kippenberg20081,treutlein20071}. 

One of the major experimental goals is to use neutral atoms to coherently manipulate the state of the
oscillator. There are several proposals aimed at achieving this by employing atoms in a cavity with a moving mirror
\cite{meiser20061,genes20081,ian20081}, or by coupling atoms by means of a reflective membrane,  where the lattice
trapping the atoms is built by reflecting a laser beam off the membrane \cite{camerer20111,vogell20131}.

Such opto-mechanical systems, composed by nano-mechanical
oscillators and atoms interfaced via optical quantum buses, have been recently discussed in the context
of quantum non-demolition Bell measurements and the ability to prepare EPR entangled states
\cite{hammerer20091}. Also ions are proposed as transducers for electromechanical oscillators
\cite{hensihger20051} while other proposals involve the coupling between oscillators and dipolar
molecules \cite{singh20081}. A growing interest, both theoretical and experimental, is therefore
apparent in the study of nanomechanical oscillators and their interaction with other quantum systems
\cite{schwab2005a,oconnell20101}. In these systems it is possible to achieve different levels of
coherent control by incorporating them into combined (hybrid) devices, involving single electron
transistors \cite{blencowe2005a,gurvitz2008a} and point contacts (PCs) \cite{mozyrsky2002a}, microwave
cavities in superconducting regime \cite{regal2008}, or superconducting qubits \cite{armour2008a,blencowe2008a}.
These numerous experiments and theoretical proposals indicate the feasibility  of studying quantum correlations,
quantum control of mechanical force sensors and decoherence in the regime where strong coherent coupling is achieved.

In particular, the experimental advances mentioned above will enhance our ability to test fundamental quantum properties,
such as decoherence in a well controlled setting \cite{hunger20101}. The determination of decoherence rates
to a  high accuracy, and their comparison to theoretical predictions will be possible in a near future. One
particularly interesting goal would be to explore the quantum-to-classical transition \cite{zurek19911} in
the dynamics of the mechanical oscillator, and the possibly anomalous decoherence that the oscillator may
exhibit when in contact with a BEC.

In \cite{brouard20111} it was analysed the dynamics of an oscillator coupled to a BEC trapped in a symmetric double well
potential, with the atomic current dependent on the oscillator coordinate. The fact that the bosons tunnel into a single
state, rather than into a broad energy zone, as in the case of a point contact, gives the decoherence process unusual
properties. Thus, a qubit monitored by a BEC undergoes an anomalously slow state-dependent decoherence \cite{sokolovski2009a},
while its decoherence in the presence of a PC is exponential in time. Similarly, a harmonic quantum oscillator whose position
is being monitored is capable of retaining some, or even all, of its coherence \cite{brouard20111}. One of the reasons for such
behaviour lies in the fact that a displaced harmonic oscillator maintains its equidistant level structure, and its motion
remains periodic even when coupled to a BEC via its position. This may not be true if there is even a small degree of
anharmonicity in the oscillator's motion. The effect of anharmonicity on the decoherence rate of an oscillator coupled
to a BEC trapped in a double well structure is the subject of this work. 
% or even entire %
%Under these circumstances the BEC is able to monitor the oscillator evolution at the cost of introducing
%decoherence to the oscillator dynamics. In particular, it was showed that an oscillator monitored by a BEC
%does not, in general, undergoes a quantum-to-classical transition \cite{mozyrsky2002a,katz2007a} and it might, in some cases,
%may retain some degree of coherence. 
%However, if there is some degree of anharmonicity in the motion of the
%oscillator these conclusions might be modified significantly thus preventing coherence even for small
%non-linearities. The aim of this work is to study the decoherence effects induced by a BEC trapped in a
%double-well structure onto an \emph{anharmonic} oscillator when both, BEC and oscillator, are coupled
%through the position of the oscillator. The effect of anharmonicity in the decoherence process is the object
%of this communication.

The rest of the paper is organized as follows. A brief description of the model is presented in Section II. A quasiclassical
analysis of the system dynamics and its implications in the appearance of decoherence is presented in Section III A.
Section III B contains results of a full quantum analysis. Our conclusions are presented in Section IV.

\section{The 'gatekeeper' model}

\begin{figure}[ht]
\centering{\includegraphics[width=12cm,height=10cm]{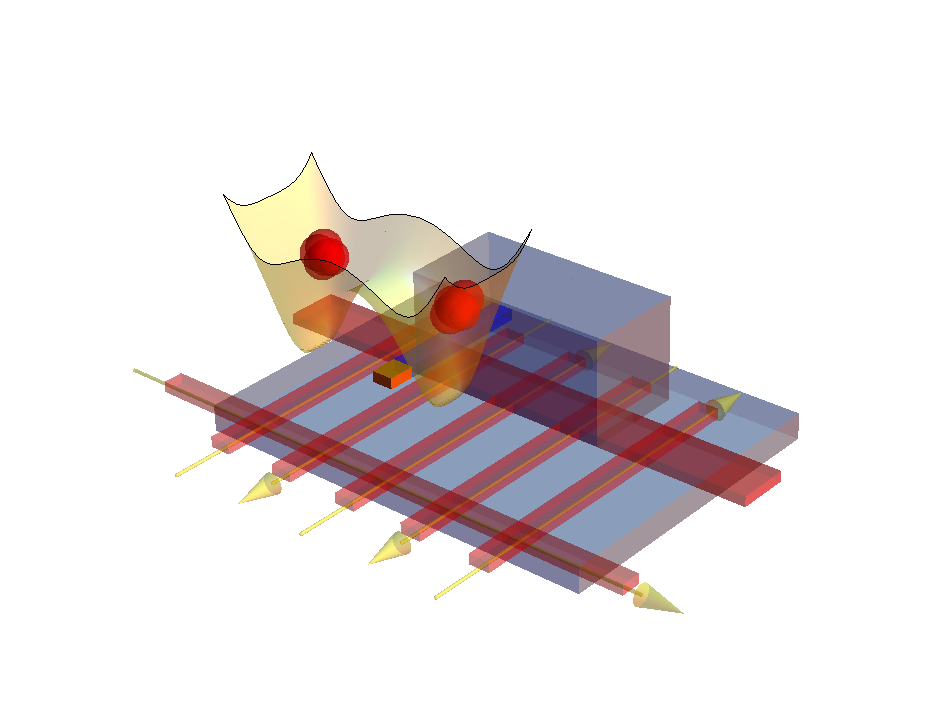}}
\caption{(Color online) Graphical representation of an atom chip that can be a realization of the model studied
in this work. An array of wires is used to create two dimple traps that contain atoms (red). The traps are separated
by tunable barriers. The yellow arrows indicate the currents used to create the traps. On top of the wire
configuration that supports the traps, the cantilever (blue) is mounted with a tip (orange) that interacts
with the atoms. The figure is inspired in the work \cite{reichel20001}.}
\label{fig:4}
\end{figure}

We are interested in an anharmonic nanomechanical oscillator coupled to a BEC in such a way that its position
influences the atomic current flowing between the wells of a double-well potential in which the BEC's atoms are
trapped. Thus we consider, in one dimension, a quartic anharmonic oscillator of mass $m$ and frequency $\omega_0$,
described by the Hamiltonian
%(we put $\hbar=1$)
%
\begin{equation}
\os H=
%&\hbar \omega_0\left(\hat a \da \hat a+\frac{1}{2}\right)+\frac{\hbar\omega_a}{4}(\hat a\da +\hat a)^4\nonumber\\
\frac{\hat P^2}{2m}+\frac{1}{2}m \omega_0^2 \hat X^2+\frac{\beta}{4}\hat X^4,
\end{equation}
where $\beta$ controls anharmonicity of the potential. The oscillator is coupled to a BEC composed by non-interacting
bosonic atoms trapped in a symmetric double well potential. Without tunnelling the atoms may occupy the left or right
well
\begin{equation}
\cn H= E_0\,(\hat c^{\dagger}_L \hat c_L+\hat c^{\dagger}_R \hat c_R),
%-\hbar\Omega_1\,(\hat c^{\dagger}_L \hat c_R+\hat c^{\dagger}_R \hat c_L),\label{eq:1}\nonumber\\
%\cn B&=&\hat c^{\dagger}_L \hat c_R+\hat c^{\dagger}_R \hat c_L.
\end{equation}
where the operator $\hat c^{\dagger}_{L} (\hat c^{\dagger}_{R})$ creates a boson in the ground state of the left (right)
well. We choose the coupling to be linear in the the BEC's tunnelling operator $\hat T=\hat c^{\dagger}_L \hat c_R+
\hat c^{\dagger}_R \hat c_L$, and assuming oscillation to be small, linearise it also in the oscillator's position $X$
\begin{equation}\label{eq:1}
 \hat H_{int}=\hbar \Omega(\hat X)\otimes\hat T
 \approx  \hbar[\Omega_0+\Omega_1\hat X] \,(\hat c^{\dagger}_L \hat c_R+\hat c^{\dagger}_R \hat c_L).
%\cn B&=&\hat c^{\dagger}_L \hat c_R+\hat c^{\dagger}_R \hat c_L.
\end{equation}
Now with tunnelling switched on, the full Hamiltonian is given by  
 %$\hat a^\dagger (\hat a)$ is the
%creation (annihilation) operator of the harmonic oscillator with mass $m$ and frequency $\omega_0$, and
%$\hbar\beta\equiv(2m\omega_0)^2\omega_a$ characterizes the anharmonicity of the potential.
%The Hamiltonian describing the dynamics of the combined system, oscillator plus condensate, is given by
\cite{sokolovski2009a}
\begin{equation}
\hat H=\os H + \cn H + \hat H_{int}.
\end{equation}
Initially the system is prepared in a product state, and its density matrix has the form 
\begin{equation}\label{INIT}
\hat \rho(0)=\os \rho(0)\otimes \cn \rho(0),
\end{equation}
where $\cn \rho(0)$ corresponds to some non-equilibrium state of the BEC. The flow of bosons across the barrier dividing
the wells depends on $\hat X$, and may be used to extract information about the oscillator's position. The back action
provided by such a measurement on the observed oscillator is the subject of this paper. A sketch of a possible experimental
setup is shown in Fig.1.
% We will choose $\Omega_1=0$ so as to exclude a constant background current \cite{sokolovski2009a}.
%\newline

Since the trap is symmetric, the tunnelling operator commutes with the $\cn H$,  $[\cn H,\hat T]=0$. Suppose the BEC is
prepared in a stationary state $|\tilde{\phi}_n\rangle$,
\begin{eqnarray}
&&|\tilde{\phi}_n \rangle=
%\frac{\left(\hat c^{\dagger}_L+\hat c^{\dagger}_R\right)^{N-n}\,
%\left(\hat c^{\dagger}_L-\hat c^{\dagger}_R\right)^n}{\sqrt{2^N (N-n)!\, n!}}\left|0 \right>_{\rm con},
[2^N (N-n)! n!]^{-1/2}
\left(\hat c^{\dagger}_L+\hat c^{\dagger}_R\right)^{N-n}
\left(\hat c^{\dagger}_L-\hat c^{\dagger}_R\right)^n|0 \rangle_{\rm con}\qquad(n=0,1,..,N),\\
&&\hat c_L|0 \rangle_{\rm con} = \hat c_R|0 \rangle_{\rm con}=0,\nonumber
\end{eqnarray}
for which we also have $\cn H|\tilde{\phi}_n\rangle =N E_0$, and $\hat T|\tilde{\phi}_n\rangle=(N-2n)|\tilde{\phi}_n \rangle$.
It is readily seen that the BEC will continue in $|\tilde{\phi}_n \rangle$, while the oscillator would experience a constant
energy shift of $\hbar \Omega_0(N-2n)$ and an additional force $\varphi_n\equiv\hbar\Omega_1(N-2 n)$. Moreover, the density
matrix of the oscillator at a time $t$, $\os\rho(t)={\rm Tr}_{\rm con}\{\rho(t)\}$, is given by a weighted sum of density matrices
$\os\rho^{(n)}$ evolved from $\os \rho(0)$ under different forces $\varphi_n$
\begin{equation}
\os\rho(t)=\sum_n P(n)\,\os\rho^{(n)}(t),
\label{eq:5}
\end{equation}
where
\begin{eqnarray}
&&\os\rho^{(n)}(t)\equiv e^{-i\os{\mathcal{H}} (\varphi_n)t/\hbar} 
\os\rho(0) e^{i\os{\mathcal{H}} (\varphi_n)t/\hbar}\\
&&\os{\mathcal{H}}(\varphi_n)\equiv \os H+ \varphi_n \hat X,\nonumber
\end{eqnarray}
and $P(n)$ is the probability to find the BEC in a state $|\tilde{\phi}_n\rangle$ at $t=0$,
\begin{equation}
P(n)={\rm Tr}_{\hbox{\scriptsize{con}}}\left[|\tilde{\phi}_{\it n}\rangle\langle\tilde{\phi}_{\it n}|\cn{\rho}(0)\right].
\label{eq:5a}
\end{equation}
(We note that the constant energy shifts cancel, and do not contribute to the oscillators evolution).
 
Similarly, for the expectation value of an oscillator's observable $\hat O$, we have
\begin{equation}
\langle \hat O \rangle= \sum_n P(n) {\rm Tr}_{\rm osc}\left[\os\rho^{(n)}(t) \hat O \right]= \sum_n P(n)
{\rm Tr}_{\rm osc}\left[\os\rho(0) \hat O_n(t) \right],
\label{eq:6}
\end{equation}
with $\hat O_n(t)=e^{i\os{\mathcal{H}} (\varphi_n)t/\hbar}\hat Oe^{-i\os{\mathcal{H}} (\varphi_n)t/\hbar}$.

We can consider a limit in which the number of bosons in the BEC increases, while the individual tunnelling probability is
reduced, so that there is a finite atomic current between the two wells, 
\begin{equation}\label{irrev}
N\to\infty, \quad \Omega_{0,1}\to 0, \quad \Omega_{0,1}\sqrt{N}=\kappa_{0,1}. 
\end{equation}
Preparing all the atoms in the left well,
\begin{equation}\label{left}
\cn\rho(0)=|\psi_0\rangle\langle\psi_0|, \quad|\psi_0 \rangle\equiv (\hat c^{\dagger}_L)^N |0
\rangle_{{\rm{con}}}/\sqrt{N!},
\end{equation}
we have a source of practically irreversible current, since the Rabi period after which the BEC returns to its initial state is
now very large \cite {sokolovski2009a}. Measuring after a time $t$ the number of atoms in the right well gives information about
the oscillator's past \cite{sokolovski2009b}. Also the sums in Eqs.(\ref{eq:5}) and (\ref{eq:6}) can be replaced by integrals,
$\sum_nP(n) \to \int d\varphi P(\varphi)$. Continuous distribution of forces corresponding to the initial state (\ref{left}) is
Gaussian,
\begin{equation}
P(\varphi)=\frac{e^{-\varphi^2/2\Delta_{\varphi}^2}}{\sqrt{2\pi \Delta_{\varphi}^2}},
\label{Pcont}
\end{equation}
with $\Delta_\varphi^2\equiv 2m \hbar\omega_0\kappa^2$ \cite {sokolovski2009a}.

\section{Monitoring position of a quartic anharmonic oscillator}

The process of decoherence appears in general in the dynamics of averages of particular observables. Then, instead
of analysing the density matrix (\ref{eq:5}), it is convenient to consider the oscillator's mean position, thus
choosing the operator $\hat O$ in Eq.(\ref{eq:6}) as the operator $\hat X$,
\[
\langle \hat X(t)\rangle=\sum_n P(n) \langle \hat X_{n}(t)\rangle,
\]
which reads, in the continuous limit,
\begin{widetext}
\begin{eqnarray}\label{equis}
&&\langle \hat X(t)\rangle=\int_{-\infty}^{\infty} d\varphi\,
P(\varphi) \left<\hat X_{\varphi}(t)
\right>\nonumber\\
&=&\int_{-\infty}^{\infty} d\varphi\, P(\varphi)\left\{
\sum_i \left<\psi^{\varphi}_i|\os\rho(0)|\psi^{\varphi}_i\right>
\left<\psi^{\varphi}_i|\hat X|\psi^{\varphi}_i\right>
+\sum_{i\ne j} e^{-i\left(E^{\varphi}_i-E^{\varphi}_j\right)t/\hbar}
\left<\psi^{\varphi}_i|\os\rho(0)|\psi^{\varphi}_j\right>
\left<\psi^{\varphi}_j|\hat X|\psi^{\varphi}_i\right>
\right\},\\
\nonumber
\end{eqnarray}
\end{widetext}
with $\os{\mathcal{H}}(\varphi)|\psi^{\varphi}_i \rangle = E^{\varphi}_i |\psi^{\varphi}_i\rangle$. Indeed, it can be
observed that in the case $E^{\varphi}_i-E_j^{\varphi}\ne const(\varphi)$, the exponentials in the second term of
Eq.(\ref{equis}) are rapidly-oscillating functions and consequently such term will vanish. Therefore $\os\rho(t)$
[and with it the averages (\ref{eq:6})] will tend to stationary values as $t\rightarrow \infty$. Without such a
cancellation, the oscillator will not be able to reach a steady state no matter how long one waits.

Equation (\ref{equis}) is the starting point of the quantum calculations we shall present.

\subsection{Time evolution in Wigner space}

%The figures of this section are generated by the code CatStateNanoMechanicalV4.nb and WignerBECV1.f 

\begin{figure}[ht]
%C:\Users\dalonso\Documents\Research\Works\Measurement_SantiDimitriGurvitz\Paper\paper2\alfa_4_beta_0p05_deltaphi_0p1_ene_100
%Generated by CatStateNanoMechanicalV4_alfa_4_beta_0p05_deltaphi_0p1_ene_100.nb
\centerline{\includegraphics[width=12cm]{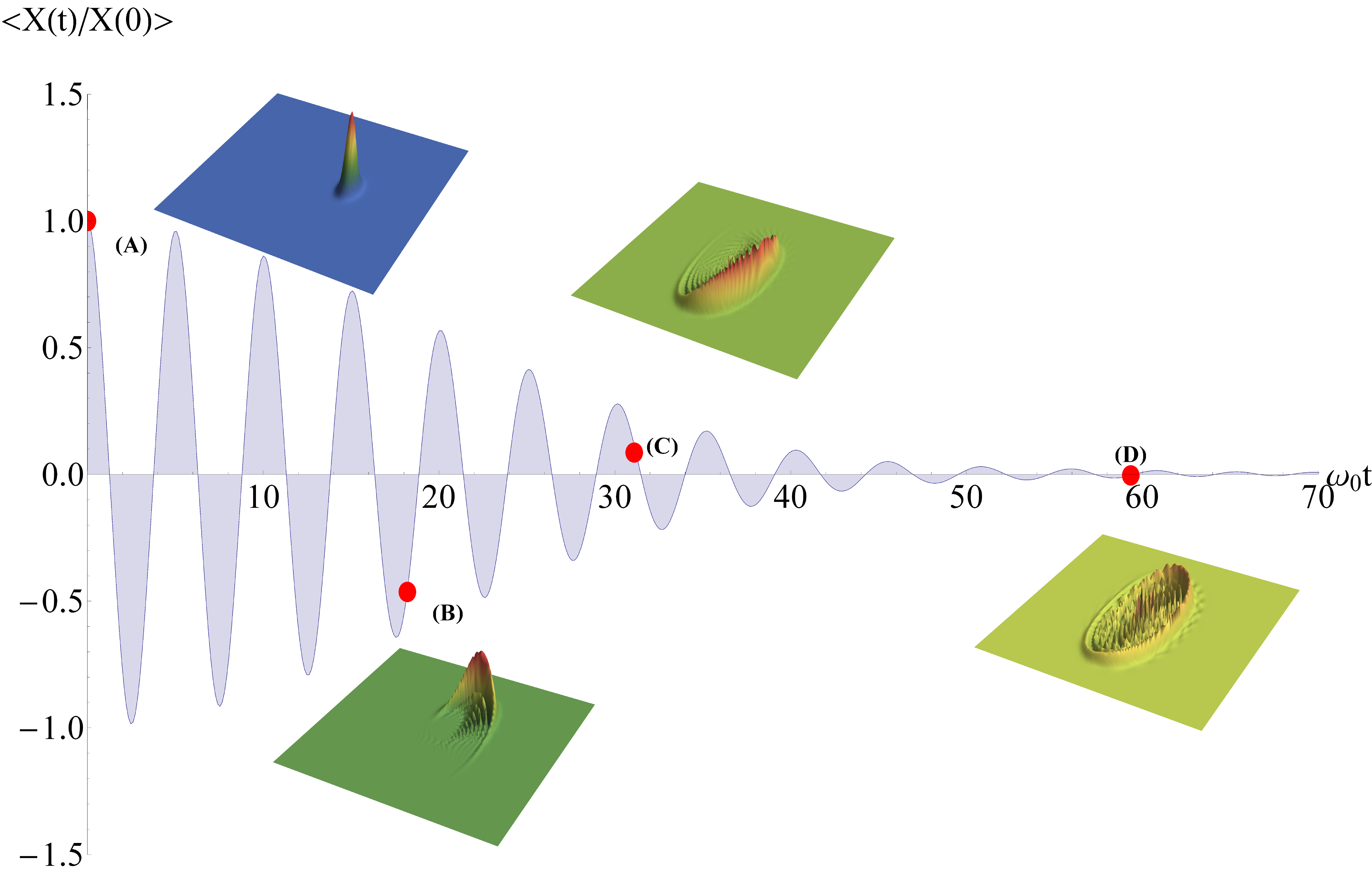}}
\caption{(Color online) Decay of the expectation value of position and some Wigner functions of the oscillator at different times.}
\label{quantumdecay}
\end{figure}

We shall first look at the time evolution of a state of the oscillator in Wigner representation when it is influenced
by the condensate. In that representation, the state develops structure as time progresses. An initial state of the
oscillator evolves according to a Schr\"{o}dinger equation that takes into account the effect of the condensate. In one
dimension the Wigner function associated to a quantum state $\hat \rho$ is defined as \cite{wigner19321}
\begin{equation}
W(x,p,t)=\frac{1}{2 \pi \hbar}\int dq \, e^{i p q/\hbar} \Big \langle x-\frac{q}{2} \,\Big|\,\hat \rho\,\Big|\,
x+\frac{q}{2} \Big \rangle.
\label{wignerfunction}
\end{equation}
It is convenient to write a state $\hat \rho$ of the oscillator for a given time $t$ in terms of the eigenstates
$|\varphi_n \rangle$ of the harmonic oscillator as
%
%\begin{equation}
%|\psi \rangle =\sum_{n=0}^{\infty} c_n(t)|\varphi_n \rangle,
%\label{eq:}
%\end{equation}
%
%in terms of which
%
\begin{equation}
\hat \rho(t)=\sum_{n,m} c_{nm}(t) |\varphi_n \rangle \langle \varphi_m |.
\label{eq:}
\end{equation}
By inserting this expression in (\ref{wignerfunction}) it follows the Wigner function corresponding to the state of the
oscillator as
\begin{equation}
W(x,p,t)=\sum_{n,m} c_{nm}(t) w_{nm}(x,p),
\label{wignerosc}
\end{equation}
where
\begin{equation}
w_{nm}(x,p)=\frac{1}{2 \pi \hbar} \int \, e^{i p q/\hbar} \Big \langle x-\frac{q}{2} \,\Big|\,\varphi_n
\Big\rangle \Big\langle \varphi_m \Big|\,  x+\frac{q}{2} \Big\rangle.
\label{eq:}
\end{equation}
The integral in this expression has a representation in terms of generalized Laguerre polynomials $L_n^a$ yielding
\begin{equation}
w_{nm}(x,p)=
\begin{cases}
\frac{(-1)^n e^{-|z|^2}}{\pi \hbar} 
\Big( \frac{n! \, 2^m}{m!\, 2^n}\Big)^{1/2} (z^*)^{m-n} L_{n}^{m-n}(2|z|^2 ),& \mbox{if} (n\le m) \\
\\
\frac{(-1)^m e^{-|z|^2}}{\pi \hbar} \Big( \frac{m! \, 2^n}{n!\, 2^m}\Big)^{1/2} (z^{})^{n-m} L_{m}^{n-m}(2|z|^2 )
& \mbox{if} (n\ge m),
\end{cases}
\label{wnm}
\end{equation}
with $z=x \sqrt{m \omega_0/\hbar}+i p\sqrt{\hbar m \omega_0}$. When equation (\ref{wnm}) is combined with
equation (\ref{wignerosc}) it follows an analytical expression for the Wigner function of the oscillator. In
Figure \ref{quantumdecay} it is plotted the Wigner function of the oscillator at different times. It can be
seen that as time proceeds an initial coherent state starts to spread over phase space developing a highly
oscillatory structure, that at the end is responsible for the decay of the position expectation value. It is
precisely the study of such decay and of its details the subject of this work. 

%\begin{figure}[ht]
%C:\Users\dalonso\Documents\Research\Works\Measurement_SantiDimitriGurvitz\Paper\paper2\alfa_4_beta_0p05_deltaphi_0p1_ene_100
%Generated by CatStateNanoMechanicalV4_alfa_4_beta_0p05_deltaphi_0p1_ene_100.nb
%\centerline{\includegraphics[width=12cm]{gr1.png}}
%\caption{Decay of the expectation value of position and some Wigner functions of the oscillator at different times.}
%\label{quantumdecay}
%\end{figure}

\subsection{Quasi classical approximation}

We begin by evaluating the quasi classical limit of (\ref{equis}) which is conveniently obtained by employing the Weyl-Wigner
representation \cite{weyl19271,wigner19321}. For the average of an arbitrary operator describing the oscillator, $\hat O_n(t)$,
we write
\begin{equation}\label{1}
\langle \hat O_n(t) \rangle = \int_{-\infty}^\infty dx\, \int_{-\infty}^\infty dp\, W(x,p)\,O_n(x,p,t),
%\label{averageWW}
\end{equation}
where $W(x,p)$ and $O_n(q,p,t)$ are the Weyl-Wigner transforms of the initial oscillator's state, and of the operator $\hat O_n(t)$
in its Heisenberg representation,  
\begin{equation}\label{2}
O_n(x,p,t)=\int dq \, e^{i p q/\hbar} \Big \langle x-\frac{q}{2} \,\Big|\,\hat O_n(t)\,\Big|\,  x+\frac{q}{2} \Big \rangle.
%\label{eq:}
\end{equation}

The dynamical aspects of the formalism are contained in the equation of motion for $O_n(x,p,t)$
\begin{equation}\label{3}
%\label{qliouville}
\partial_t O_n(x,p,t)=\{\{ \mathcal{H}(x,p;\varphi_n),O_n(x,p,t)\}\},
%\frac{2}{\hbar} \mathcal{H}(x,p;\varphi_n) \, \hbox{{\normalsize S}in}{\big( \frac{\hbar}{2}
%\stackrel{\leftrightarrow}{\Lambda}\big)} \, O_n(x,p,t).
\end{equation}
where $\mathcal{H}(x,p;\varphi_n)$ is the Wigner representation of the quantum Hamiltonian operator $\os{\mathcal{H}}$, and
the Moyal (Sine) bracket is defined, as usual \cite{Moyal}, by $$\{\{f(x,p),g(x,p)\}\}\equiv \frac{2}{\hbar} f(x,p)
\sin[\frac{\hbar}{2}(\stackrel{\leftarrow}{\partial_p}\stackrel{\rightarrow}{\partial_x}-\stackrel{\leftarrow}{\partial_x}
\stackrel{\rightarrow}{\partial_p})]g(x,p)$$ with, e.g.,  $f(x,p)\stackrel{\leftarrow}{\partial_p}\stackrel{\rightarrow}
{\partial_x}g(x,p)\equiv \partial_p f(x,p) \partial_x g(x,p)$.
%$(\varphi_n)$ and $f(x,p) \stackrel{\leftrightarrow}{\Lambda} g(x,p):= f(x,p)\stackrel{\leftarrow}{\partial_p}
%\stackrel{\rightarrow}{\partial_x}g(x,p)-f(x,p)\stackrel{\leftarrow}{\partial_x} \stackrel{\rightarrow}
%{\partial_p}g(x,p)\equiv \partial_p f(x,p) \partial_x g(x,p)-\partial_x f(x,p) \partial_p g(x,p)$.
%>From equation (\ref{qliouville}) it is possible to obtain exact solutions in some cases. Nonetheless the
%computation of approximate solutions is viable if we consider a formal expansion in the Planck's constant
Equations (\ref{1}) - (\ref{3}) are a convenient starting point for our quasi classical analysis. Expanding $O_n(x,p,t)=
X_n(x,p,t)$ in powers of $\hbar$, as
\begin{equation}
\label{qclass}
X_n(x,p,t)=X^{cl}_n(x,p,t)+ \hbar^2 X^{(2)}_n(x,p,t)+{\cal{O}}(\hbar^4),
\end{equation}
and recalling that as $\hbar \to 0$ the Moyal bracket reduces to the classical Poisson bracket $\left \{ *,* \right \}$
\cite{gaspard1995a}, we obtain an approximate equation of motion for $X^{cl}_n(x,p,t)$,
%
%\begin{equation}\label{POISS}
%\partial_t O^{cl}_n(x,p,t)= \Big\{\mathcal{H}(x,p;\varphi_n),O^{cl}_n(x,p,t) \Big\}.
%\end{equation}
%
%Equation (\ref{POISS}) must be solved subject to an appropriate initial condition, and for our choice $ \hat O_n= \hat X_n$, we have
%
\begin{equation}\label{POISS1}
\partial_t X^{cl}_n(x,p,t)= \Big\{\mathcal{H}(x,p;\varphi_n),X^{cl}_n(x,p,t) \Big\}, 
\end{equation}
with the initial condition
\begin{equation}\label{POISS3}
X^{cl}_n(x,p,t=0)= x, 
\end{equation}
which is obtained by evaluating the integral (\ref{2}) for $t=0$. A solution of the classical equation of motion (\ref{POISS1}) is
any function $\mathcal{F}(x_t(x,p),p_t(x,p))$, provided $x_t$ and $p_t$ satisfy the Hamiltonian equations of motion (a dot denotes
the time derivative)
\begin{eqnarray}\label{class}
\dot {p}_t&=-\partial_{x_t}\mathcal{H}(x_t,p_t;\varphi_n) =&-m \omega^2_0 x_t-\beta x^3_t-\varphi_n \nonumber\\
\dot{x}_t&=\partial_{p_t}\mathcal{H}(x_t,p_t;\varphi_n) =&p_t/m,
\end{eqnarray}
subject to $x_{t=0}=x$ and $p_{t=0}=p$. With the help of the initial condition (\ref{POISS3}) we identify $\mathcal{F}(x,p)$ with $x$,
so that $X^{cl}_n(x,p,t)=x_t(x,p)$. Thus, $X^{cl}_n(x,p,t)$ is just the position, at a time $t$, of the oscillator whose initial
position and momentum at $t=0$ were $x$ and $p$, respectively, 

It is readily seen that in the quasi classical limit, the task of calculating the mean oscillator's position at time $t$, 
\begin{equation}
\langle \hat X(t) \rangle_{qcl}=\sum_n P(n)\left [\int_{-\infty}^\infty dx\, \int_{-\infty}^\infty dp\, W(x,p)\,X^{cl}_n(x,p,t)\right],
\label{quasiclass}
\end{equation}
reduces to choosing initial phase space distribution $W(x,p)$, which contains all quantum effects, and evaluating classical oscillator
trajectories for different values of the induced force $\varphi_n$. In the limit of small anharmonicity ($\frac{\beta x_0^2}{\omega_0^2}
\ll 1$) the system will be shown to behave as a set of harmonic oscillators each one with a slightly shifted frequency and the semiclassical approximation described by the previous equation is very accurate for the purpose of studying the decoherence effect induced by the coupling with the BEC system. To support the accuracy of the approximations made comparisons with the results obtained by exact numerical integration of the Schr\"odinger equation are provided. The numerics give strong support to the semiclassical approximation in the limit we are
considering.

\subsection{A coherent initial state. Small anharmonicity}

Next we specify our analysis to the case where the oscillator is prepared in a coherent state, whose Weyl-Wigner transform is given by
\begin{equation}
\label{wigner}
W(x,p)=\frac{1}{\pi\hbar}e^{-\frac{m \omega_0}{\hbar}(x-x_0)^2-\frac{1}{m\hbar\omega_0}(p-p_0)^2}.
\end{equation}
There are no analytical solutions for a classical anharmonic oscillator. However, the decoherence effects absent for a harmonic oscillator,
appear already in the limit of small anharmonicity ($\frac{\beta x_0^2}{\omega_0^2}\ll 1$). In this case approximate oscillator trajectories
can be obtained, e.g., by the method of strained coordinates (Lindsted-Poincar\'e method) for periodic solutions \cite{kevorkian1981a},
which we will describe here briefly. We begin by considering a trajectory such that at some $t=t_0$ it passes through some $x^0$ with a
zero momentum, $x_{t_0}=x^0$, $p_{t_0}=0$. This can be represented by a sum of harmonic functions with phases and amplitudes modified at
different orders in $\beta$. An approximate solution to the first order in $\beta$ for the phase and to zero order for the amplitude is
given by \cite{kevorkian1981a,marinca2012}
\begin{equation}
\label{xclass}
X^{cl}_n({x}^0,0,t)=-\frac{\varphi_n}{m \omega^2_0}+\left(\frac{\varphi_n}{m \omega^2_0}+{x}^0\right)\cos\left\{
\omega_0\left[1+\beta \Delta({x}^0,0;\varphi_n)\right](t-t_0)\right\},
\end{equation}
with
\begin{equation}
\Delta(x,p;\varphi_n)=\frac{3}
{4(m \omega_0^2)^2}\,{\cal{H}}_0(x,p;\varphi_n)+\frac{15}{8 m^3 \omega_0^6}\,\varphi_n^2+{\cal{O}(\beta)},\qquad\qquad\qquad\qquad
\end{equation}
and
\begin{equation}
\label{H0}
{\cal{H}}_0(x,p;\varphi_n)=\frac{p^2}{2 m}+\frac{1}{2}m \omega_0^2 x^2+\varphi_n x.\qquad\qquad\qquad\qquad\qquad\qquad\qquad\qquad
\end{equation}

The solution corresponding to an arbitrary choice of initial $x$ and $p$ is then obtained 
 %at the same order in $\beta$, can be
%computed consistently from the previous solution
by choosing in Eq.(\ref{xclass}) ${x}^0$ and $t_0$ in such a way that the trajectory 
specified by Eq.(\ref{xclass}) would, at $t=0$, pass through $x$ with the desired momentum $p$.
Explicitly, we have
\begin{equation}
\label{xclass2}
X_n^{cl}(x,p,t)= -\frac{\varphi_n}{m \omega^2_0}+\left(\frac{\varphi_n}{m\omega^2_0}+x\right)\cos(\omega_1 t )
+\frac{p}{m \omega_0} \sin(\omega_1 t ),
\end{equation}
which describes a harmonic motion whose frequency is modified both by the anharmonicity
of the oscillator potential and the presence of the BEC, and also depends on the initial position 
$x$ and momentum $p$ of the oscilllator, 
$$\omega_1=\omega_0 \left[1 +\beta \Delta(x,p;\varphi_n)\right].$$
% yet we will not include  them
%since the decoherence, which is the main subject of this paper, appears already at this level of accuracy.
Higher order corrections in $\beta$ can be systematically obtained if necessary, although decoherence for small enough $\beta$
($\frac{\beta x_0^2}{\omega_0^2}\ll 1$) is accurately described at this level of approximation. In particular a third harmonic
contribution, with an amplitude which is first order in $\beta$, is negligible as compared to the decoherence/dephasing effect
we will show next to be produced by the combined effect of frequency shift in the first harmonic (which is also first order
in $\beta$) and the interaction with the BEC, which implies a superposition of signals with frequencies
$\omega_1=\omega_0 \left[1 +\beta \Delta(x,p;\varphi_n)\right]$ that contain terms $\beta\varphi_n$ and $\beta\varphi_n^2$,
which is at the end the origin of the decoherence effect we illustrate in this work (see the details in the Appendix).

Replacing in Eq.(\ref{quasiclass}) the summation over discrete levels of the BEC by integration %over $\varphi$ 
as described in Sect. II (and changing the discrete subscript $n$ to a continuos index $\varphi$) yields
\begin{eqnarray}
\label{xsm}
\langle \hat X(t) \rangle_{qcl}= \frac{1}{\sqrt{2\pi\Delta_{\varphi}^2}}\int_{-\infty}^\infty\int_{-\infty}^\infty
\int_{-\infty}^\infty  d\varphi\,
%\int_{-\infty}^\infty 
dx\,
 %\int_{-\infty}^\infty 
dp\, \exp(-\varphi^2/2\Delta_{\varphi}^2)
\, W(x,p)\, X^{cl}_\varphi(x,p,t).
\end{eqnarray}
The integral in Eq.(\ref{xsm}), with $W(x,p)$ and $X^{cl}_\varphi(x,p,t)$ given by Eqs.(\ref{wigner}) and (\ref{xclass2})
respectively, can be evaluated analytically, e.g.,  by formally introducing a Gaussian generating function
${\cal Z}({\bf J})$,
\begin{eqnarray}\label{Generating}
	{\cal Z}({\bf J})&=&e^{-C} \int \frac{d{\bf z}}{\pi\hbar\sqrt{2 \pi \Delta_{\varphi}^2}}
	e^{-\frac{1}{2}{\bf z}^T \cdot {\mathsf A} \cdot {\bf z}+({\mathbf B}+{\bf J})^T \cdot {\bf z}} 
	\\ \nonumber
	&=&\frac{2 e^{-C}}{\hbar\Delta_{\varphi}\sqrt{{\hbox{det}} {\mathsf A}}}\, e^{\frac{1}{2}({{\mathbf B}+\bf J})^T
        \cdot {\mathsf A}^{-1}\cdot({{\mathbf B}+\bf J})},
\end{eqnarray}
with ${\bf z}=(x,p,\varphi)$, ${\bf J}=(J_1,J_2,J_3)$,
\begin{equation}
{\mathsf A}=\left(
\begin{array}{lll}
 \frac{2 m \omega_0 }{\hbar}-\frac{3 i t \beta }{4 m \omega_0} & 0
   & -\frac{3 i t \beta }{4 m^2 \omega_0^3} \\
 0 & \frac{2}{m\hbar\omega_0}-\frac{3 i t \beta }{4 m^3 \omega_0^3} & 0 \nonumber \\
 -\frac{3 i t \beta }{4 m^2 \omega_0^3} & 0 & 
\frac{1}{\Delta_\varphi^2}-\frac{15 i t \beta }{4 m^3 \omega_0^5}
\end{array}
\right),\qquad\qquad\quad
\end{equation}
\begin{equation}
{\mathbf B}=\left(
\begin{array}{l}
\frac{2 m \omega_0  x_0}{\hbar}\\
\frac{2 p_0}{m \omega_0\hbar}\\
0
\end{array}
\right),\,{\hbox{and}}\quad
C=-i\omega_0 t+\frac{2}{\hbar\omega_0}\left(\frac{p_0^2}{2 m}+\frac{1}{2}m \omega_0^2  x_0^2\right). 
\end{equation}
Then defining 
%The generating function allows us evaluate the averages of the type
 $\overline{z_i}(t) \equiv\frac{\partial {\cal{Z}}}{\partial J_i}|_{{\bf J}=0}$, $i=1,2,3$,
we have
\begin{equation}\label{result}
\langle \hat X(t) \rangle_{qcl}=\frac{{\rm Re}\left[\overline {\varphi}(t)\right]}{m \omega_0^2}+
{\rm Re}\left[\overline { x}(t) \right]+\frac{{\rm Im}\left[\overline { p}(t)\right]}{m \omega_0},
\end{equation}
from which an explicit analytical expression can be derived, although we will not cite it here.
%
%where $\langle z_i(t) \rangle\equiv\frac{\partial {\cal{Z}}}{\partial J_i}|_{{\bf J}=0}$. 

Equations (\ref{xsm})-(\ref{result}) are the main result of this Section. In Fig. \ref{fig:4} we compare the analytical
results in Eq. (\ref{result}) with those obtained for $\langle \hat X(t) \rangle$ in equation (\ref{equis}) by numerical
diagonalisation of each $\os{\cal{H}}(\varphi_n)$. At $t=0$, the oscillator is prepared  in a coherent state with
$x_0=3\;\hbox{a.u.}$, and $p_0=0$, $\beta$ has been set to $0.05\;\hbox{a.u.}$ and $\omega_0=1.3\;\hbox{a.u.}$ which justifies
the perturbative approach of Eqs.(\ref{xclass})-(\ref{xclass2}).
%Atomic units are used in this work. 
The agreement between both results is good, and we proceed to use the quasi classical Eq.(\ref{result}) in order to
characterise the decoherence in the short and the long time limits.

\begin{figure}[ht]
%C:\Users\dalonso\Documents\Research\Works\Measurement_SantiDimitriGurvitz\Generating_Function.nb
\centering{\includegraphics[width=10cm]{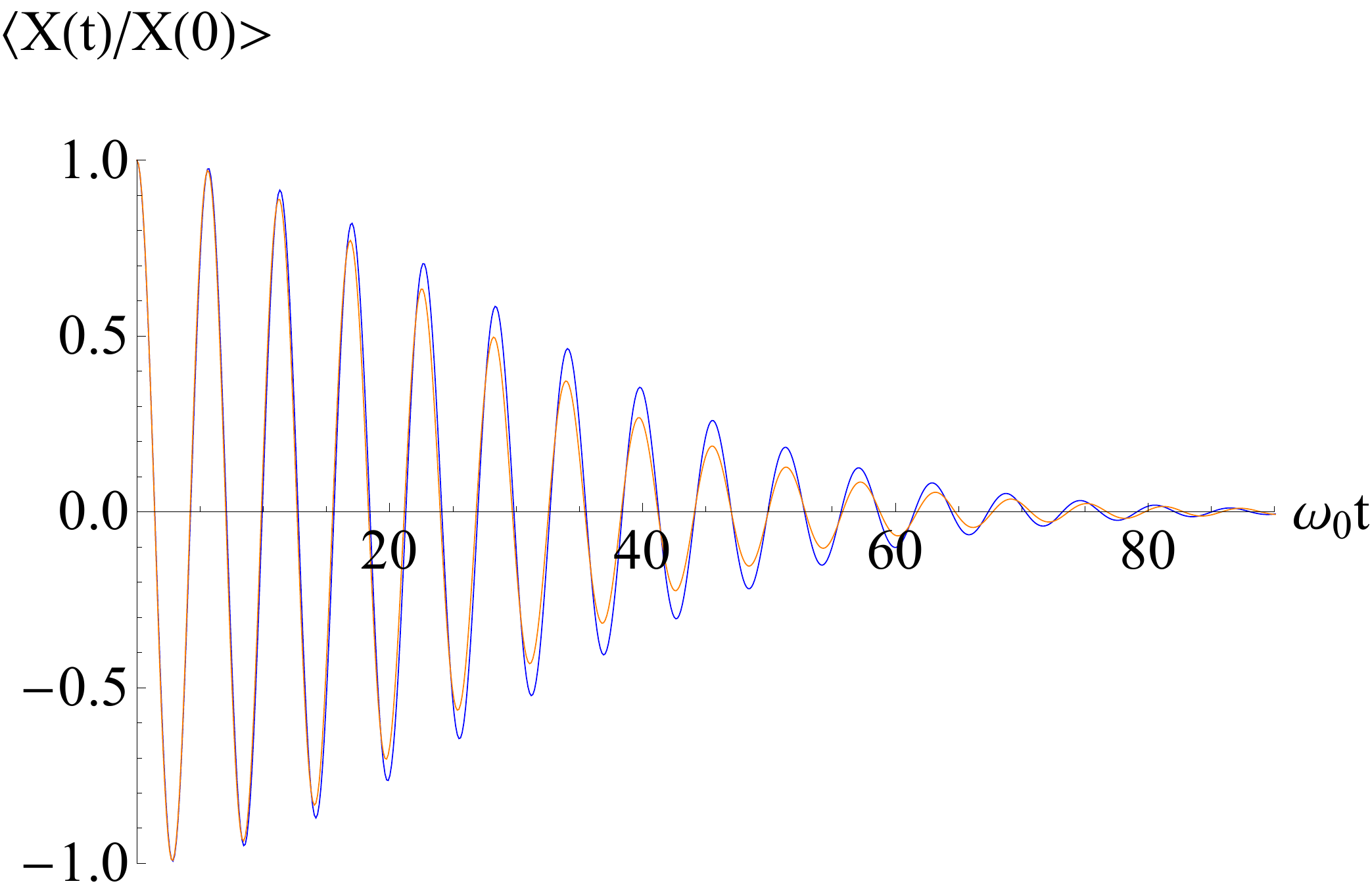}}
\caption{(Color online) Comparison between the quasiclassical solution (\ref{result}) (orange line) and the exact numerical
solution of the equations of motion (blue line) in the perturbative regime ($\beta x_0^2/2 m \omega_0^2=0.133$) for an initial
coherent state with $x_0=3$, $p_0=0$, $\omega_0=1.3$, $\beta=0.05$ and $\Delta_{\varphi}=0.1$ (atomic units are used).}
\label{fig:4}
\end{figure}

\subsection{Time scale analysis}

If the oscillator is not coupled to the condensate then it will show a coherent motion in which coherences will remain in
time leading to recurrences in the oscillator dynamics. The action of the condensate quenches such recurrences. We shall
assume that the coupling between the condensate and the oscillator is such that recurrences in the dynamics of the oscillator
have been suppressed and therefore the decoherence process occurs in a time scale much shorter than the dynamical recurrence
time of the free (uncoupled) oscillator.

There are two relevant processes with regard to the time development and decay of $\langle \hat X(t) \rangle_{qcl}$: The non-linearity
in the potential, and the interaction with the condensate. As already emphasized, when such non-linearity does not exists the coupling
between the oscillator and the condensate will not lead to a decay in oscillator's expectation values, even if variances and higher order
fluctuations are affected by such coupling \cite{brouard20111}. However, the non-linear potential together with the oscillator-condensate
interaction induces decoherence. A natural time scale linked to such non-linearity can be defined as $t_{\beta}=(3 \beta \hbar/4 m^2
\omega_0^2)^{-1}$. In addition, the interaction between the oscillator and the condensate introduces a different time scale, given by
$t_{\varphi}=(3 \beta \Delta_{\varphi}^2/4 m^3 \omega_0^5)^{-1}$. Both time scales are different and they are useful to understand the time
development of oscillator's observables. We shall analyze the dynamics in two different situations, for $t$ either smaller or larger
than both characteristic time scales.

\subsubsection{Case 1: $t\ll t_{\beta},t_{\varphi}$}

In this case the solution is accurately represented by
\begin{equation}
\langle \hat X(t) \rangle_{qcl} \simeq e^{-t^2/2 t^2_G} \left[
x_0 \cos(\omega_1 t)+\frac{p_0}{m \omega_0} \sin(\omega_1 t)\right],
\end{equation}
with $\omega_1=\omega_0+\frac{3 \beta \hbar} {4 m^2 \omega_0^2}\frac{{\cal{H}}(x_0,p_0;0)}{\hbar\omega_0}=\omega_0+\Delta \omega_0$, and
\begin{equation}
t_G=t_{\beta}\left( \frac{\hbar \omega_0 }{{\cal{H}}(x_0,p_0;0)+m \omega_0^2 x_0^2t_{\beta}/t_{\varphi}}\right)^{1/2}.
\end{equation}
%

%\begin{figure}[ht]
%%C:\Users\dalonso\Documents\Research\Works\Measurement_SantiDimitriGurvitz\Generating_Function.nb
%\centering{\includegraphics[width=10cm]{gc_2.pdf}}
%\caption{Long time behaviour: Comparison between the quasiclassical solution (\ref{result}) (orange line) and the exact numerical solution of
%the equations of motion (blue line) in the perturbative regime ($\beta x_0^2/2 m \omega_0^2 = 0.133$) for an initial coherent
%state with $x_0=3$, $p_0=0$, $\omega_0=1.3$, $\beta=0.05$ and $\Delta_{\varphi}=0.1$ (atomic units are used).}
%\label{fig:4b}
%\end{figure}
It is apparent that $t_G$ is a natural time scale associated with a Gaussian decoherence process taking place for times smaller than
$t_{\beta}$ and $t_{\varphi}$. Furthermore, a fully Gaussian decay, including a Gaussian tail, will develop if $t_G\ll t_{\beta}$,
$t_{\varphi}$. That situation will appear in a fully semiclassical regime. In the quantum domain $t_G$ will be of the order of or smaller
than $t_{\varphi}$ leading to a decay which will not be Gaussian. In the limit of harmonic potential a coherent motion as reported in
\cite{brouard20111} is recovered, emphasizing the relevance of the non-linearity to the decoherence process. 

\subsubsection{Case 2: $t\gg t_{\beta},t_{\varphi}$}
\label{sec:LongTimeBehavior}

In this case the solution can be represented by using a long time approximation by
\begin{equation}
\langle \hat X(t) \rangle_{qc} \simeq e^{{-2\frac{{\cal{H}}(x_0,p_0;0)}{\hbar\omega_0}}}
\frac{m^{11/2} \omega_0^{13/2}}{\hbar^2 \Delta \varphi} \frac{64}{9 \sqrt{3}}(\beta t)^{-5/2}
\left(
x_0 \cos \omega_0 t+\frac{p_0}{m \omega_0} \sin \omega_0 t\right)+{\cal{O}}\left({(\beta t)}^{-7/2}
\right).
\end{equation}
The agreement of this expression with the numerical exact solutions to the quantum equations has been checked for small
$\beta$ ($\frac{\beta x_0^2}{\omega_0^2}\ll 1$). We observe that if time is much larger than $t_{\beta}$ and $t_{\varphi}$
the decay is algebraic. Notably the amplitude is exponentially damped as $\exp(-{\cal{H}}(x_0,p_0;0)/\hbar \omega_0)$,
therefore this suggests that deeply within the quasiclassical regime, in consistency with the results of the previous
section, it would be difficult to observe such power law decay, indicating that such dynamics would be observable mainly
within the quantum domain. 

The decay of observables shows an initial Gaussian decay for short times and a power law decay for longer times. Our
results are consistent with those in \cite{braun20011,strunz20031,znidaric20051} where Gaussian decoherence was studied.

In the next section we present numerical simulations as well as some analytical approximate expressions describing the
dynamics of the system in the full quantum regime.

\subsection{Quantum Analysis}

For the quantum case and an arbitrary value of $\beta$ there are no analytical solutions. Numerical solutions can be
obtained efficiently by diagonalizing $\os H+\varphi \hat X$ over a truncated basis of stationary states of the
harmonic oscillator centered at the origin, for each value of $\varphi$. Convergence with respect to the number of
basis states used for the truncated diagonalization is checked for each particular value of $\beta$ and for the
initial state of the oscillator. These numerical results are used all throughout the paper to compare with
the different analytical approximations described.

%The same general discussion about the decay of the amplitude of the oscillating terms of the reduced density matrix,
%presented at the beginning of this section, can be applied to $\langle X(t)\rangle$. We are interested in
%characterizing its decay with time.

In the general case, for arbitrary values of $\beta$, many different time-dependent terms will contribute to the
sum in (\ref{equis}). However, in some limiting cases and under some approximations, analytical expressions can be
obtained that describe the time decay with accuracy.

We will consider here as an illustration the case of an initial state of the oscillator involving only a few lower-energy
states of the oscillator with frequency $\omega_0$ and a value of $\hbar\beta \ll m^2 \omega_0^3$.

Let us focus on the time-dependent part of Eq.(\ref{equis}),
\begin{equation}
\langle \hat X(t)\rangle-\langle \hat X\rangle^{st}=\frac{1}{\sqrt{2\pi\Delta_\varphi^2}}\int_{-\infty}^{\infty} d\varphi\,
e^{-\varphi^2/2\Delta_\varphi^2}\sum_{i<j} e^{-i\left(E^{\varphi}_i-E^{\varphi}_j\right)t/\hbar}F_{i,j}(\varphi)+c.c.,
\end{equation}
where the superscript ``st'' stands for stationary and with $F_{i,j}(\varphi)\equiv\left<\psi^{\varphi}_i|\os\rho(0)|
\psi^{\varphi}_j\right>\left<\psi^{\varphi}_j|\hat X|\psi^{\varphi}_i\right>
$.

Firstly, for not too large values of $\Delta_\varphi$, it is enough to write the energy eigenvalues as a second
order expansion in $\varphi$, $E^{\varphi}_i(\beta)=E^{\varphi=0}_i(\beta)+\gamma_i \varphi^2$, (the term linear in
$\varphi$ is zero because the anharmonicity is even in the $x-$coordinate). A first order approximation in $\beta$
for $E^{\varphi=0}_i(\beta)$ and $\gamma_i$,
\begin{eqnarray}
&&E^{\varphi=0}_i=\hbar\omega_0\left(i+\frac{1}{2}\right)+\frac{3\hbar^2\beta\,\left(i^2+i+\frac{1}{2}\right)}
{8m^2\omega_0^2},\\
&&\gamma_i=-\frac{1}{2 m \omega_0^2}+\frac{3\hbar\beta\left(2 i+1\right)}{4m^3\omega_0^5},
\end{eqnarray}
will be sufficient for the states that will contribute to the summation.

%With respect to the factor that depends on $\beta_i$, a better (numerical) estimation for $\beta_i-\beta_j$ is needed
%in the exponential to obtain the solution with accuracy, although a first order  approximation in $\beta$ for this
%quantity, $\beta_i-\beta_j\simeq\frac{3\hbar^2\beta\, (i-j)}{m^2\omega_0^4}$, already leads to a qualitatively good
%estimation of $\left<\wh X(t)\right>$, see below. 

A valid approximation for $F_{i,j}(\varphi)$ can be obtained to zeroth order in $\beta$. Writting the initial state of
the oscillator in terms of the stationary states of the harmonic oscillator, $\left|\psi_n\right>$, as $\os\rho(0)=
|\psi(0)\rangle\langle\psi(0)|$, with $|\psi(0)\rangle=\sum_n c_n |\psi_n\rangle$, and evaluating $\left<\psi^{\varphi}_i
|\psi_n\right>$ and $\left<\psi^{\varphi}_j|\hat X|\psi^{\varphi}_i\right>$ as integrals over $x$, one obtains
%%
%\[
%\left<x|\psi_n^{\varphi}\right>=
%\sqrt{\frac{1}{2^n\,n!}}\left(\frac{m\omega_0}{\pi\hbar}\right)^{1/4}\,H_n\left[\sqrt{\frac{m\omega_0}{\hbar}}
%\left(x-\frac{\varphi}{m\omega_0^2}\right)\right] e^{-\frac{m\omega_0}{2\hbar}\left(x-\frac{\varphi}{m\omega_0^2}
%\right)^2}
%\]
%%

%\[
\begin{equation}
\left<x|\psi_n^{\varphi}\right>=
\sqrt{\frac{1}{2^n\,n!}}\left(\frac{m\omega_0}{\pi\hbar}\right)^{1/4}\,H_n(y) e^{-y^2/2},
\end{equation}
with $y=\sqrt{\frac{m\omega_0}{\hbar}}\left(x-\frac{\varphi}{m\omega_0^2}\right)$, for the displaced $n$th stationary
state of the harmonic oscillator. Thus finally $F_{i,j}(\varphi)$ reads 
\begin{equation}
F_{i,j}(\varphi)=G_{i,j}(\varphi) e^{-\varphi^2/(2m\hbar\omega_0^3)},
%\sum_{n,m} C^*_n C_m \left<\psi_n|\psi^{\varphi}_j\right>\left<\psi^{\varphi}_j|X|\psi^{\varphi}_i\right>
%\left<\psi^{\varphi}_i|\psi_m\right>
\end{equation}
where $G_{i,j}(\varphi)=g_{i,j}^{(0)}+g_{i,j}^{(2)}\varphi^2$ is a known polynomial of $\varphi$. The odd
powers will not contribute to the integral. Furthermore, only terms up to the second order will be kept.

\begin{figure}[ht]
%\centerline{\includegraphics[width=10cm]{quantum_decay.pdf}}
\centerline{\includegraphics[width=10cm]{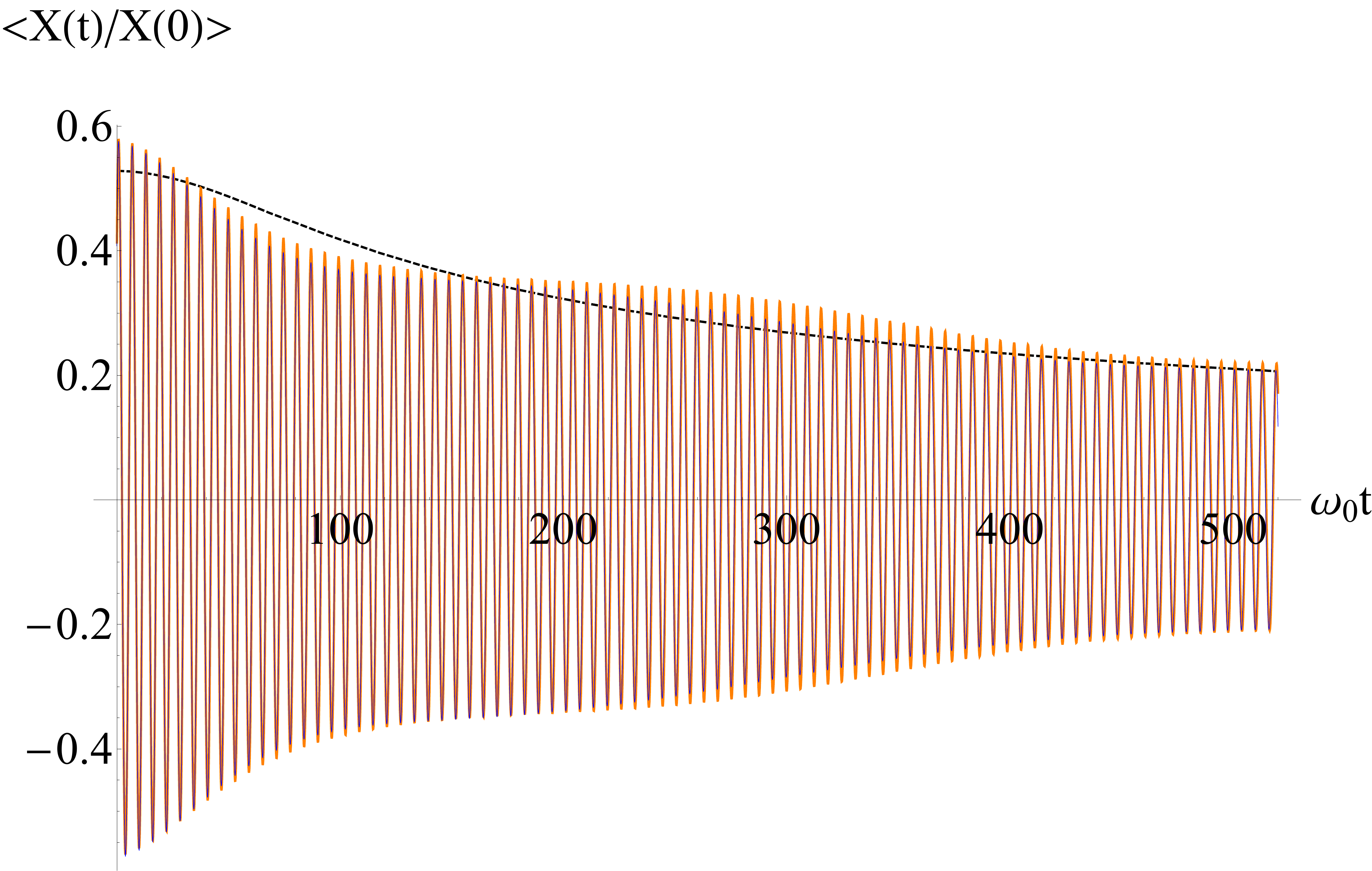}}
\caption{(Color online) Mean value of position as a function of $\omega_0 t$ (normalized to its initial value) for an initial state
of the oscillator $|\psi(0)\rangle=\left(1+i\right)|\psi_0\rangle/\sqrt{3}+i|\psi_1\rangle/\sqrt{3}$. Numerical result
(solid line) and the analytical approximation given by Eq. (\ref{aprox1}) (dashed line) are almost indistinguishable
in the figure. Frequency and amplitude of the signal are described with great accuracy by the analytical expression.
The envelope of the analytical approximation in Eq. (\ref{aprox0}) is shown for reference (dotted line).
The inset shows the details of the evolution for the expectation value of position in the large time region.
$\beta=0.05$, $\omega_0=1.3$, $\Delta_\varphi=0.7$ (a.u.)}
\label{numdecay}
\end{figure}

For an initial state being a combination of the first two lower-energy states of the harmonic oscillator,
$|\psi(0)\rangle=c_0 |\psi_0\rangle+c_1|\psi_1\rangle$, the mean value of position then reads,
\begin{eqnarray}\label{aprox1}
\langle\hat X(t)\rangle=\langle \hat X\rangle^{st}&+&\sum_{n=0,1} \frac{e^{-i\omega_{n,n+1}t}}{\sqrt{2\Delta_{\varphi}^2}}
\left(g_{n,n+1}^{(0)}\sqrt{\frac{1}{1/(2\Delta_{\varphi}^2)+1/(2m\hbar\omega_0^3)+i(\gamma_n-\gamma_{n+1})t}}\right.\nonumber\\
&+&\left.\frac{1}{2} g_{n,n+1}^{(2)} \left(\frac{1}{1/(2\Delta_{\varphi}^2)+1/(2m\hbar\omega_0^3)+i(\gamma_n-\gamma_{n+1})t}
\right)^{3/2}\right)+{\rm c.c},
\end{eqnarray}
where $\omega_{n,n+1}\equiv E_n^{\varphi=0}-E_{n+1}^{\varphi=0}$. Fig. \ref{numdecay} shows exact numerical 
results compared to the analytical approximation given by Eq. (\ref{aprox1}). The inset presents the two curves for
large time, where the analytical approximation is shown to reproduce both the frequency and the amplitude of the
oscillations with great accuracy. The coefficients $g_{n,n+1}^{(0)}$ and $g_{n,n+1}^{(2)}$ depend on the coefficients
in the Hermite polynomials, $H_n(x)$, as well as on $c_0$ and $c_1$ characterizing the state $|\psi(0)\rangle$,
\begin{eqnarray}
&&g_{0,1}^{(0)}=\left(\frac{\hbar}{2m\omega_0}\right)^{1/2}\,c_0 c_1^*\nonumber\\
&&g_{0,1}^{(2)}=-\left(\frac{2m\omega_0}{\hbar}\right)^{1/2}\,\frac{c_0^* c_1+c_0 c_1^*}{4m^2\omega_0^4}\nonumber\\
&&g_{1,2}^{(0)}=0\nonumber\\
&&g_{1,2}^{(2)}=\left(\frac{2m\omega_0}{\hbar}\right)^{1/2}\,\frac{c_0^* c_1+2c_0 c_1^*}{4m^2\omega_0^4}.\nonumber
\end{eqnarray}
%
%Numerical values for these coefficients are given in the figure caption (ATTENTION WHERE ARE THEY?).
If only the term $n=0$ is considered, a first order approximation in $\beta$ is used for the difference $\gamma_0-
\gamma_1=-3\hbar\beta/(2m^3\omega_0^5)$, and $G_{0,1}(\varphi)$ is taken as $g_{0,1}^{(0)}$. A good qualitative
approximation is already obtained for this case,
\begin{equation}\label{aprox0}
\langle\hat X(t)\rangle\simeq\langle\hat X\rangle^{st}+g_{0,1}^{(0)}\,e^{-i\omega_{0,1}t} \sqrt{\frac{1}{1+2
\Delta_\varphi^2/(2m\hbar\omega_0^3)+3i\Delta_\varphi^2\frac{\hbar\beta}{m^3\omega_0^5}t}}+c.c.
\end{equation}
The envelope of $\langle\hat X(t)\rangle$ in Eq. (\ref{aprox0}) is also shown for comparison in Fig \ref{numdecay}.
Power law decay for the amplitude of the oscillations is observed, the different powers that contribute to the
result depending on the initial state, being of the general form $t^{-k/2}$, with $k$ being an integer.

\section{Conclusions}

A detailed study of decoherence of an anharmonic oscillator in contact with a BEC trapped in a double well potential
is performed. The oscillator is coupled to the BEC through its position. In contrast with the harmonic oscillator case,
for which coherent behaviour has been reported, the anharmonic oscillator presents anomalous decoherence (non-exponential).
In the quasiclassical domain there are two clearly distinguishable regimes. For short times, decoherence appears to be
Gaussian with a well defined time scale. Such time scale depends on the degree of anharmonicity of the oscillator as well
as on the energy distribution of the initial state of the whole system. The higher the anharmonicity of the oscillator
and/or the energy distribution of the initial state, the faster is the decay of coherence in short time scales. All that
in consistency with a quantum-to-classical transition. On the other hand, at long times coherence decays algebraically.
The particular power of the decay is characteristic of the initial states considered. The observation of both time regimes
requires a very fine tuning of the initial state of the system, in particular of its initial energy distribution as
measured by the energy variance. In the full quantum domain decoherence manifests itself in a combination of pure algebraic
decay processes for all times of the form $t^{-k/2}$ ($k$ an integer) according to the decomposition of the initial state of
the oscillator in the harmonic oscillator number basis. This would allow to observe coherent motion for longer times that
what it would be possible with an exponential decay. Our results show that a slight anharmonicity in the confining potential
of the oscillator is sufficient to observe anomalous decoherence. 

\begin{acknowledgments}
We are grateful to Shmuel Gurvitz for useful discussions. D.A. thanks the warm hospitality of the Max Planck Institute for
the Physics of Complex Systems at Dresden where part of this work was completed. Two of us (D.A. and S.B.) acknowledge
financial support provided by Spanish MICINN (Grant No. FIS2010-19998) and the European Union (FEDER). \\
\end{acknowledgments}

\appendix*

\section{Third harmonic contribution}

It is known that the contribution of higher harmonics to the dynamics of the anharmonic oscillator may be relevant \cite{marinca2012}.
In fact, one can explicitly compute the solution $x(t)$ of $\ddot{x}=-\omega_0^2 x-\beta x^3$ with $x(0)=x_0$
and $\dot x(0)=0$, which includes those higher harmonics i.e.
\begin{eqnarray}
x(t)&&=x_0 \cos \omega t+\frac{\beta x_0^3}{32 \omega^2}\left( \cos 3 \omega t
-\cos \omega t \right)+\frac{\beta^2 x_0^5}{1024 \omega^4}\left( \cos 5\omega t
-\cos \omega t \right)+..., 
\label{eq:1} \\
\hbox{with}\quad \omega^2&&=\frac{1}{16} \left[6 \beta  x_0^2+8 \omega_0^2+\sqrt{30 \beta ^2 x_0^4+96
\beta  x_0^2 \omega_0^2+64 \omega_0^4}\right].\nonumber
\end{eqnarray}

Such solution is rather accurate even for a non moderate anharmonic contribution.
However, in this work we restrict ourselves to the analysis of small anharmonicity for which
$\frac{\beta x_0^2}{\omega_0^2}\ll 1$ and hence $\omega\approx \omega_0+
\frac{3\beta x_0^2}{8 \omega_0}$. In such limit the phases are slightly corrected
and the amplitudes of higher harmonics are negligible with respect to the first harmonic
amplitude because $\frac{\beta x_0^2}{\omega_0^2}\ll 1$ implies
$\frac{\beta x_0^2}{\omega^2}\ll 1$ and $\frac{\beta^2 x_0^4}{\omega^4}\ll 1$. So, in
the limit we are considering, the main contribution to $x(t)$ comes from the first
harmonic. Let us remark that as soon as the anharmonicity starts to be important higher
harmonics should be considered but this is out of our aim in the present work.

At this stage a second aspect becomes relevant. The whole quantum signal (or its semiclassical
approximation) is a superposition of individual $x(t;x_0,p_0,\varphi)$ that are averaged over
initial conditions $x_0,p_0$ and over the parameter $\varphi$ that measures the action of the
condensate on the oscillator, see equations (\ref{xclass})-(\ref{H0}). The phase correction
depends on the initial condition and $\varphi$. Therefore, when averaging, the net result of
such superposition will be a dephased signal, that eventually decays. It happens that the decay
will be shown to be fast enough so that along the decay time, $x(t;x_0,p_0,\varphi)$, described
by its first harmonic approximation, will be in fact an accurate description of the oscillators
dynamics. The conclusion is that to characterize the decay in the parameter domain we are
studying ($\frac{\beta x_0^2}{\omega_0^2}\ll 1$), it is enough to take into account the
first correction in the phase and the first harmonic approximation.

%\bibliography{C:/Users/dalonso/Documents/Research/Bibliography/References4_25022013} 
%

\end{document}